\documentclass[apj]{emulateapj}

\input{epsf}

\begin{document}
\def\lax    {\ifmmode{_<\atop^{\sim}}\else{${_<\atop^{\sim}}$}\fi}
\def\gax    {\ifmmode{_>\atop^{\sim}}\else{${_>\atop^{\sim}}$}\fi}
\def\gtorder{\mathrel{\raise.3ex\hbox{$>$}\mkern-14mu
             \lower0.6ex\hbox{$\sim$}}}
\def\ltorder{\mathrel{\raise.3ex\hbox{$<$}\mkern-14mu
             \lower0.6ex\hbox{$\sim$}}}
 

 
\title{XMM-Newton discovery of 217 s pulsations in the brightest persistent supersoft X-ray source in M31} 

\author{Sergey P. Trudolyubov\altaffilmark{1} and William C. Priedhorsky\altaffilmark{2}} 

\altaffiltext{1}{Institute of Geophysics and Planetary Physics, University of California, Riverside, CA 92521}

\altaffiltext{2}{Los Alamos National Laboratory, Los Alamos, NM 87545}

\begin{abstract}
We report on the discovery of a periodic modulation in the bright supersoft X-ray source XMMU J004252.5+411540 
detected in the 2000-2004 {\em XMM-Newton} observations of M31. The source exhibits X-ray pulsations with a period 
P$\sim$217.7 s and a quasi-sinusoidal pulse shape and pulsed fraction $\sim$7-11\%. We did not detect statistically 
significant changes in the pulsation period on the time scale of 4 years. The X-ray spectra of XMMU J004252.5+411540 
are extremely soft and can be approximated with an absorbed blackbody of temperature 62-77 eV and a weak power law 
tail of photon index $\Gamma\sim$1.7-3.1 in the 0.2-3.0 keV energy band. The X-ray properties of the source and 
the absence of an optical/UV counterpart brighter than 19 mag suggest that it belongs to M31. The estimated bolometric 
luminosity of the source varies between $\sim$2$\times10^{38}$ and $\sim$8$\times10^{38}$ ergs s$^{-1}$ at 760 kpc, 
depending on the choice of spectral model. The X-ray pulsations and supersoft spectrum of XMMU J004252.5+411540 imply 
that it is almost certainly an accreting white dwarf, steadily burning hydrogen-rich material on its surface. We 
interpret X-ray pulsations as a signature of the strong magnetic field of the rotating white dwarf. Assuming that the 
X-ray source is powered by disk accretion, we estimate its surface field strength to be in the range $4\times10^{5}$ G 
$<B_{0}<8\times10^{6}$ G. XMMU J004252.5+411540 is the second supersoft X-ray source in M31 showing coherent pulsations, 
after the transient supersoft source XMMU J004319.4+411758 with 865.5 s pulsation period. 
\end{abstract}

\keywords{galaxies: individual (M31) --- novae, cataclysmic variables --- X-rays: binaries --- X-rays: stars} 

\section{INTRODUCTION}
Luminous supersoft X-ray sources (SSSs) (Kahabka \& van den Heuvel 2006, and references therein) were first discovered 
in the Magellanic Clouds with the {\em Einstein} observatory and later were established as a major new source class 
based on the results of {\em ROSAT} observations. SSSs have very soft spectra typically described by blackbody models 
with temperatures of $\sim 20-80$ eV with no strong hard component, and luminosities of $\sim 10^{35}-10^{38}$ ergs 
s$^{-1}$. Although supersoft X-ray sources are not a homogeneous class, the observed properties of the majority of 
SSSs are consistent with those of accreting white dwarfs (WD) in binary systems that are steadily or cyclically 
burning hydrogen-rich matter (van den Heuvel et al. 1992). The required accretion rates in these systems can be as 
high as $10^{-7}$ M$_{\odot}$ year$^{-1}$. Another subclass of SSSs are single highly evolved stars on their way to 
WD phase. In addition, a number of more luminous ($L_{\rm X} \sim 10^{38}-10^{40}$ ergs s$^{-1}$) X-ray sources also 
classified as SSS have been recently discovered in nearby galaxies, with some of them proposed as intermediate mass 
black hole (IMBH) candidates (Fabbiano 2006, and references therein). 

The nearby, giant spiral M31 presents an excellent opportunity to study various X-ray source populations. Earlier 
observations of M31 with the {\em ROSAT} satellite revealed a significant population of supersoft X-ray sources 
(Supper et al. 1997, 2001; Kahabka 1999; Greiner 2000). The advent of a new generation of X-ray telescopes ({\em Chandra} 
and {\em XMM-Newton}) has allowed us to study SSSs in a much greater detail (Osborne et al. 2001; Trudolyubov et al. 
2001, 2005; DiStefano et al. 2004; Greiner et al. 2004; Orio 2006). Most of the SSS detected in M31 appear to be 
transient/recurrent or highly variable in X-rays. The observations with {\em XMM-Newton} and {\em Chandra} are starting 
to provide valuable information on the short-term variability of M31 SSSs: timing studies of M31 X-ray sources have led 
to the discovery of 865 s pulsations in the transient SSS XMMU J004319.4+411758 \cite{O01}, and significant short-term 
variations in other SSSs (Orio 2006; Trudolyubov, Priedhorsky \& C\'ordova 2007).  

The X-ray source XMMU J004252.5+411540 was discovered in the M31 field by the {\em Einstein} observatory (source \#69 
in Trinchieri \& Fabbiano (1991)) and detected in subsequent observations with {\em ROSAT} (source \#58 in Primini, 
Forman \& Jones (1993)), {\em Chandra} (source r2-12 in Kong et al. (2002)) and {\em XMM-Newton} (source \#352 in Pietsch 
et al. (2005)). The {\em Chandra} and {\em XMM-Newton} spectroscopy of the source revealed a thermal spectrum with effective 
temperature of $\sim 60-70$ eV (DiStefano et al. 2004; Orio 2006; Trudolyubov, Priedhorsky \& C\'ordova 2007), placing it 
in the supersoft source class. Assuming the distance of 760 kpc (van den Bergh 2000), the estimated unabsorbed luminosity 
of J004252.5+411540 in the 0.3-1.5 keV energy band can be as high as $\sim$5$\times10^{38}$ ergs s$^{-1}$, making it 
the brightest persistent supersoft source in M31.

In this paper, we report on the discovery of the coherent 217.7 s pulsations in the flux of XMMU J004252.5+411540, using 
the archival data of {\em XMM-Newton} observations. We also study X-ray spectral properties of the source, search for 
its optical/UV counterparts and discuss its nature.  

\section{OBSERVATIONS AND DATA REDUCTION}
In our analysis we used the data of four longest available 2000-2004 {\em XMM-Newton} observations of the central region 
of M31 (Table \ref{obslog}) with three European Photon Imaging Camera (EPIC) instruments (MOS1, MOS2 and pn) 
\cite{Turner01,Strueder01}, and the Optical Monitor (OM) telescope (Mason et al. 2001). We reduced {\em XMM} data using 
{\em XMM-Newton} Science Analysis System (SAS v 7.0.0)\footnote{See http://xmm.vilspa.esa.es/user}. We performed standard 
screening of the original X-ray data to exclude time intervals with high background levels, applying an upper count rate 
threshold of 20\% above average background level. The standard SAS tool {\em barycen} was used to perform barycentric 
correction on the original EPIC event files used for timing analysis.

We generated EPIC-pn and MOS images of the source field in the 0.3-7.0 keV energy band, and used the SAS standard maximum 
likelihood (ML) source detection script {\em edetect\_chain} to detect point sources. We used bright X-ray sources with 
known counterparts from 2MASS and USNO-B catalogs \cite{Cutri03,Monet03} and {\em Chandra} source lists to correct EPIC 
image astrometry. The astrometric correction was also applied to the OM images, using cross-correlation with 2MASS survey 
catalog. After correction, we estimate residual systematic error in the source positions to be of the order $0.5 - 1\arcsec$ 
for both EPIC and OM. 

To generate EPIC-MOS source lightcurves and spectra, we used a circular region of $18\arcsec$ radius centered at the 
position of XMMU J004252.5+411540. Due to the source proximity to the edge of EPIC-pn CCD in observations $\#\#$ 1,2,3 
the source counts were extracted from the elliptical regions with semi-axes of $18\arcsec$ and $15\arcsec$, including 
more than $\sim 70\%$ of the source energy flux. The adjacent source-free regions were used to extract background spectra 
and lightcurves. The source and background spectra were then renormalized by ratio of the detector areas. In this analysis 
we use valid pn events with pattern 0-4 (single and double) and pattern 0-12 (single-quadruple) events for MOS cameras. 
Because of the extreme softness of the source X-ray spectrum, only data in the 0.2-3 keV energy range was used in the 
spectral analysis. For timing analysis, we used data in the $0.2 - 1$ keV  energy band to maximize sensitivity. To 
synchronize both source and background lightcurves from individual EPIC detectors, we used the identical time filtering 
criteria based on Mission Relative Time (MRT), following the procedure described in Barnard et al. (2007). The background 
lightcurves were not subtracted from the source lightcurves, but were used later to estimate the background contribution 
in the calculation of the source pulsed fractions.

The energy spectra were grouped to contain a minimum of 25 counts per spectral bin in order to allow $\chi^{2}$ statistics, 
and fit to analytic models using the XSPEC v.12\footnote{http://heasarc.gsfc.nasa.gov/docs/xanadu/xspec/index.html} fitting 
package \cite{arnaud96}. MOS1 and MOS2 data were fitted simultaneously, but with normalizations varying independently. For 
timing analysis we used standard XANADU/XRONOS v.5\footnote{http://heasarc.gsfc.nasa.gov/docs/xanadu/xronos/xronos.html} 
tasks.

In the following analysis we assume M31 distance of 760 kpc (van den Bergh 2000). All parameter errors quoted are 68\% 
($1\sigma$) confidence limits.

\section{RESULTS}
\begin{figure}
\includegraphics[clip=true,scale=0.55]{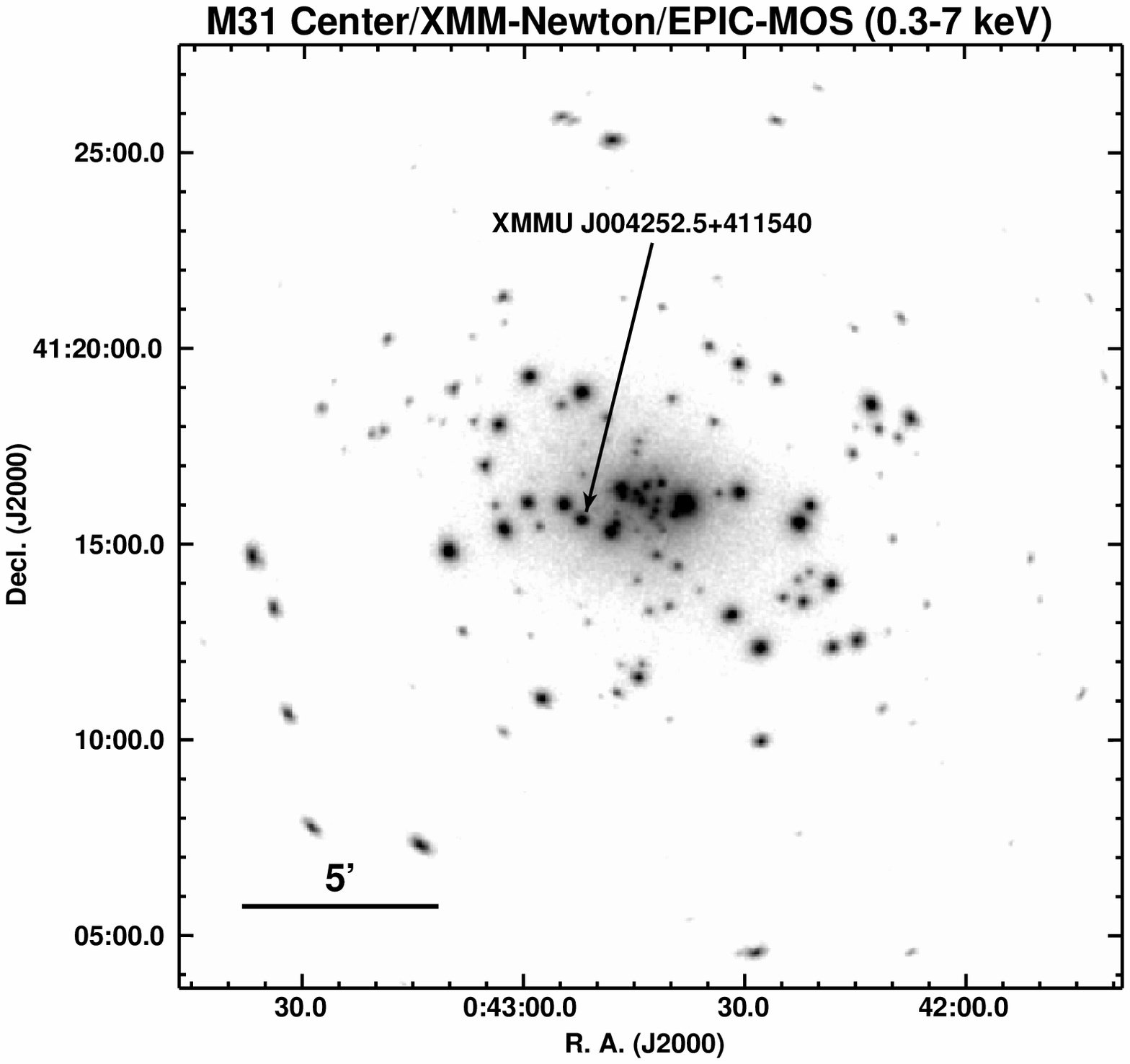}\includegraphics[clip=true,scale=0.55]{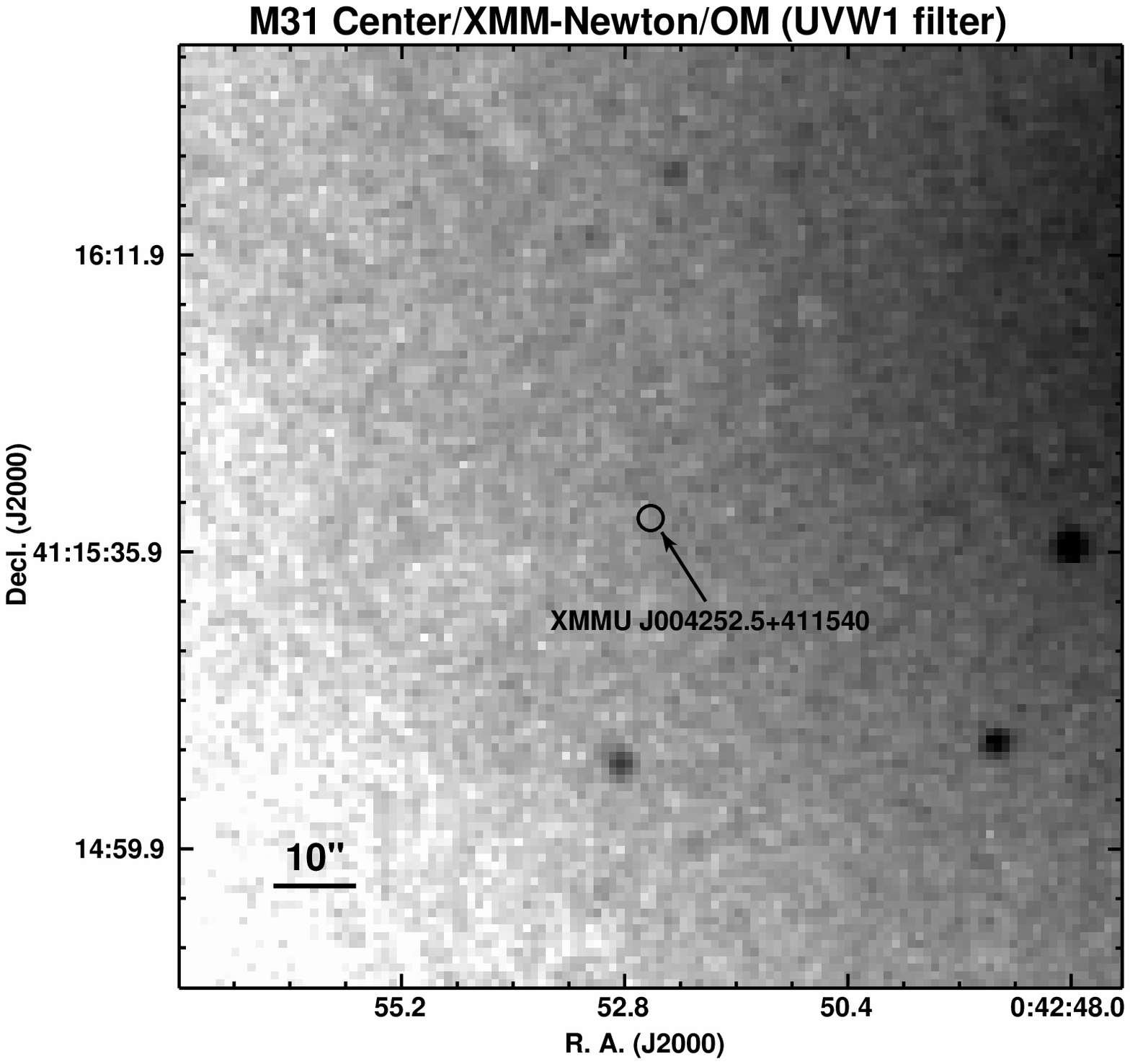}\\
\caption{{\em Left:} Combined 0.3-7 keV {\em XMM}/EPIC-MOS image covering central region of M31. The position of supersoft pulsating source XMMU J004252.5+411540 is marked with an arrow. {\em Right:} XMM-Newton/OM UVW1 (291 nm) band image of M31 field taken on Jan. 6, 2002. The image is a 2$\arcmin\times2\arcmin$ square centered on the XMMU J004252.5+411540 position. The localization of XMMU J004252.5+411540 is shown with black circle of $1.5\arcsec$ radius ($3\sigma$).\label{image_general}}
\end{figure}

Based on the {\em Chandra} aspect solution, limited by $\sim0.5\arcsec$ systematics, the source location is 
$\alpha = 00^{h} 42^{m} 52.52^{s}$, $\delta = 41\arcdeg 15\arcmin 40\arcsec$ (2000 equinox) \cite{Kong02,VG07}(Fig. 
\ref{image_general}, {\em left}). We used the data of {\em XMM-Newton}/OM observations to search for UV counterparts 
to the source. We did not detect any stellar-like objects in the OM images down to the limit of $\sim19^{m}$ in the 
OM UVW1 (291 nm) band within the error circle of XMMU J004252.5+411540 (Fig. \ref{image_general}, {\em right}).

\subsection{X-ray Pulsations}
\begin{figure}
\epsscale{1.10}
\plotone{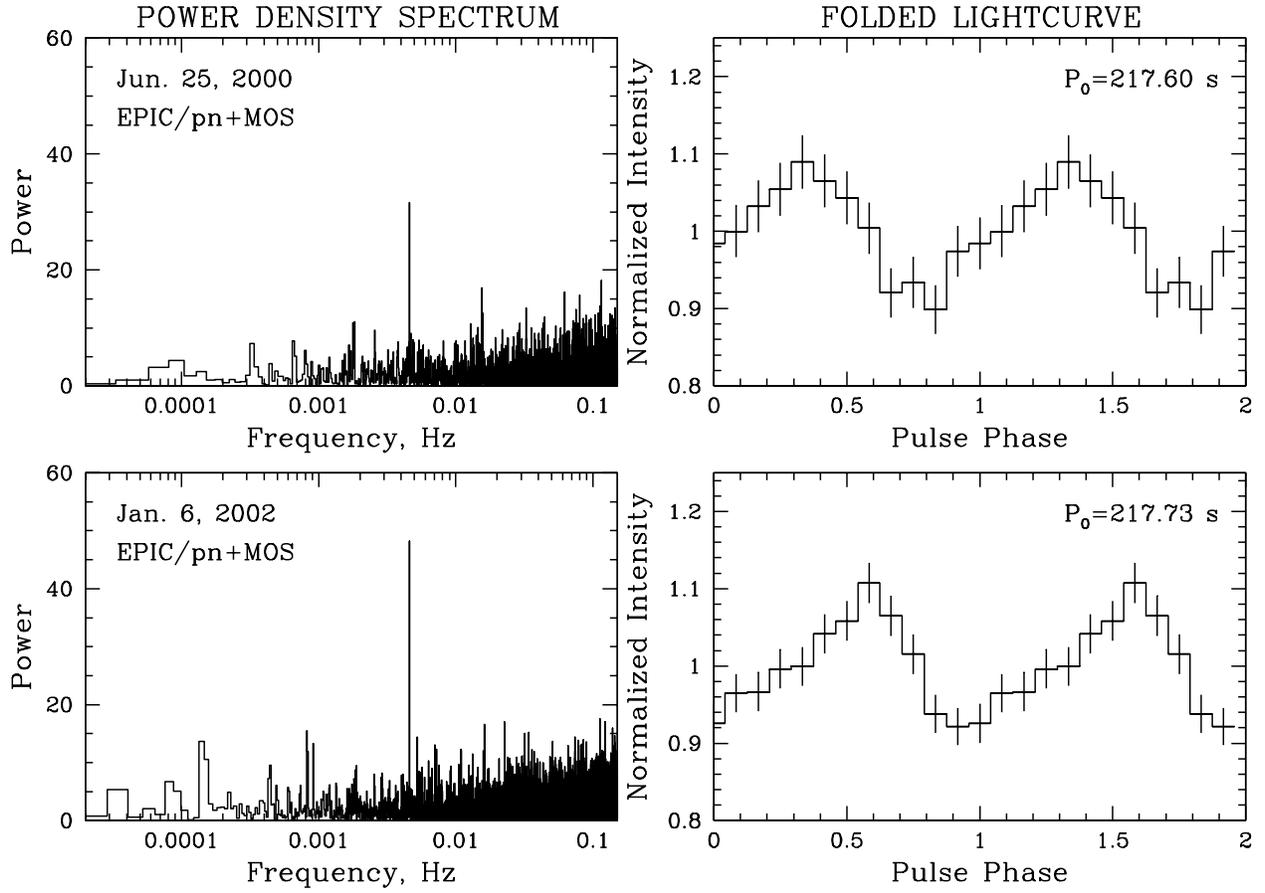}
\caption{(Left) Power spectra of XMMU J004252.5+411540 obtained using the data of 2000 Jun. 25 (upper panel) and 2002 
Jan. 6 {\em XMM-Newton}/EPIC observations (lower panel) in the 0.2-1 keV energy band. (Right) Corresponding pulse 
profiles folded with most likely pulsation periods, corrected for background.\label{pds_efold}}
\end{figure}

We performed timing analysis of XMMU J004252.5+411540 using the data from all three {\em XMM-Newton}/EPIC detectors in 
the 0.2-1 keV energy band. After a barycentric correction of the photon arrival times in the original event lists, we 
performed a Fast Fourier Transform (FFT) analysis using standard XRONOS task {\em powspec}, in order to search for 
coherent periodicities. For the analysis of {\em XMM-Newton} data, we used combined synchronized EPIC-pn and MOS 
lightcurves with 2.6 s time bins to improve sensitivity. We found strong peaks in the Fourier spectra of data from the 
2000 Jun. 25 and 2002 Jan. 6 {\em XMM-Newton} observations at the frequency of $\sim$4.6$\times10^{-3}$ Hz (Fig. 
\ref{pds_efold}, {\em left panels}). The strengths of the peaks in the individual Fourier spectra (Fig. \ref{pds_efold}) 
correspond to the period detection confidence of $\sim 1.5\times10^{-3}$ and $\sim 8\times10^{-7}$ for these observations 
\cite{Vaughan94}.  

To estimate the pulsation periods more precisely, we used an epoch folding technique, assuming no period change during 
individual observations. The most likely values of the pulsation period (Table \ref{timing_spec_par}) were obtained 
fitting the peaks in the $\chi^{2}$ versus trial period distribution with a Gaussian. The period errors in Table 
\ref{timing_spec_par} were computed following the procedure described in Leahy (1987). Then the source lightcurves 
were folded using the periods determined from epoch folding analysis. The resulting folded lightcurves of the source 
in the 0.2-1 keV energy band during 2000 Jun. 25 and 2002 Jan. 6 observations are shown in Fig. \ref{pds_efold} 
({\em right panels}). Although the power density spectra of the 2001 Jun 29 and 2004 Jul. 16 observations have a much 
weaker (with excess power $\lesssim$20) peaks at the pulsation frequency, we still performed a search for the pulsations 
in the data of these observation using epoch folding technique. The epoch folding analysis of the source lightcurves 
obtained during 2001 Jun 29 and 2004 Jul. 16 observations revealed distinct peaks in the $\chi^{2}$ versus trial period 
distribution at the periods of 217.66 s and 217.75 s (Table \ref{timing_spec_par}). We did not detect a significant 
change of the pulsation period between the four observations used in the analysis. 

The source demonstrates quasi-sinusoidal pulse profiles in the 0.2-1 keV energy band during all four observations. The 
pulsed fraction, defined as (I$_{\rm max}$-I$_{\rm min}$)/(I$_{\rm max}$+I$_{\rm min}$), where I$_{\rm max}$ and 
I$_{\rm min}$ represent source intensities at the maximum and minimum of the pulse profile excluding background photons, 
is somewhat lower during 2001 Jun. 29 observation (6.7$\pm$1.2\%) when compared to the three other observations with 
pulsed fractions of 9.1-10.9\% (Table \ref{timing_spec_par}).

\begin{figure}
\epsscale{1.00}
\plotone{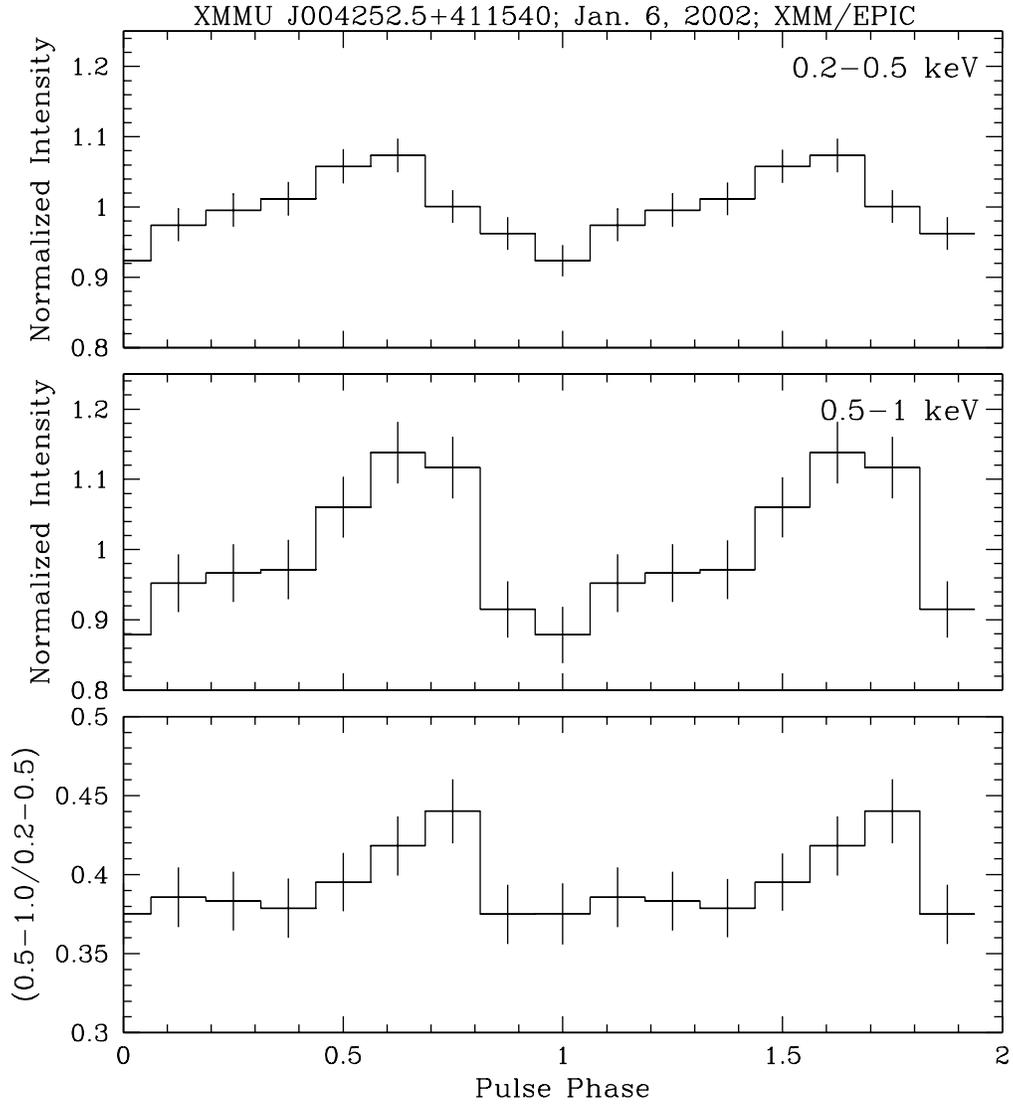}
\caption{Background-corrected normalized X-ray lightcurves of XMMU J004252.5+411540 during the 2002 Jan. 6 observation 
folded at the corresponding best period (Table \ref{timing_spec_par}) in the 0.2-0.5 and 0.5-1 keV energy bands 
({\em upper and middle panels}) along with their hardness ratio ({\em bottom panel}). 
\label{mod_energy_depend}}
\end{figure}

To investigate energy dependence of the source pulse profile, we extracted light curves in the soft (0.2-0.5 keV) and 
hard (0.5-1 keV) bands for 2002 Jan. 6 observation, and folded them at the corresponding best pulsation period (Figure 
\ref{mod_energy_depend}). As can be seen from Fig. \ref{mod_energy_depend}, there is a significant difference between the 
pulse profiles at low and high energies. The peak of the pulse profile in the 0.5-1 keV band is shifted compared to that 
in the 0.2-0.5 keV band. In addition, the pulsed fraction in the 0.5-1 keV band (12.8$\pm$2.1\%) appears to be higher 
than that at lower energies (7.5$\pm$1.2\%).

\subsection{Variability on the Time Scale of Hours}
The source is also variable on a time scales of hours during all {\em XMM-Newton} observations. The characteristic 
timescale of the variability varies between $\sim$6000 s and $\sim$8000 s. This variability can be either aperiodic 
or quasi-periodic: the existing observations are not long enough to make a definitive conclusion on its nature. The 
analysis of the long, 2001 Oct. 5 {\em Chandra}/ACIS-S observation reveals somewhat similar variability on a time 
scale of $\sim$14000 s.

\subsection{X-ray Spectra}
\begin{figure}
\epsscale{1.0}
\plotone{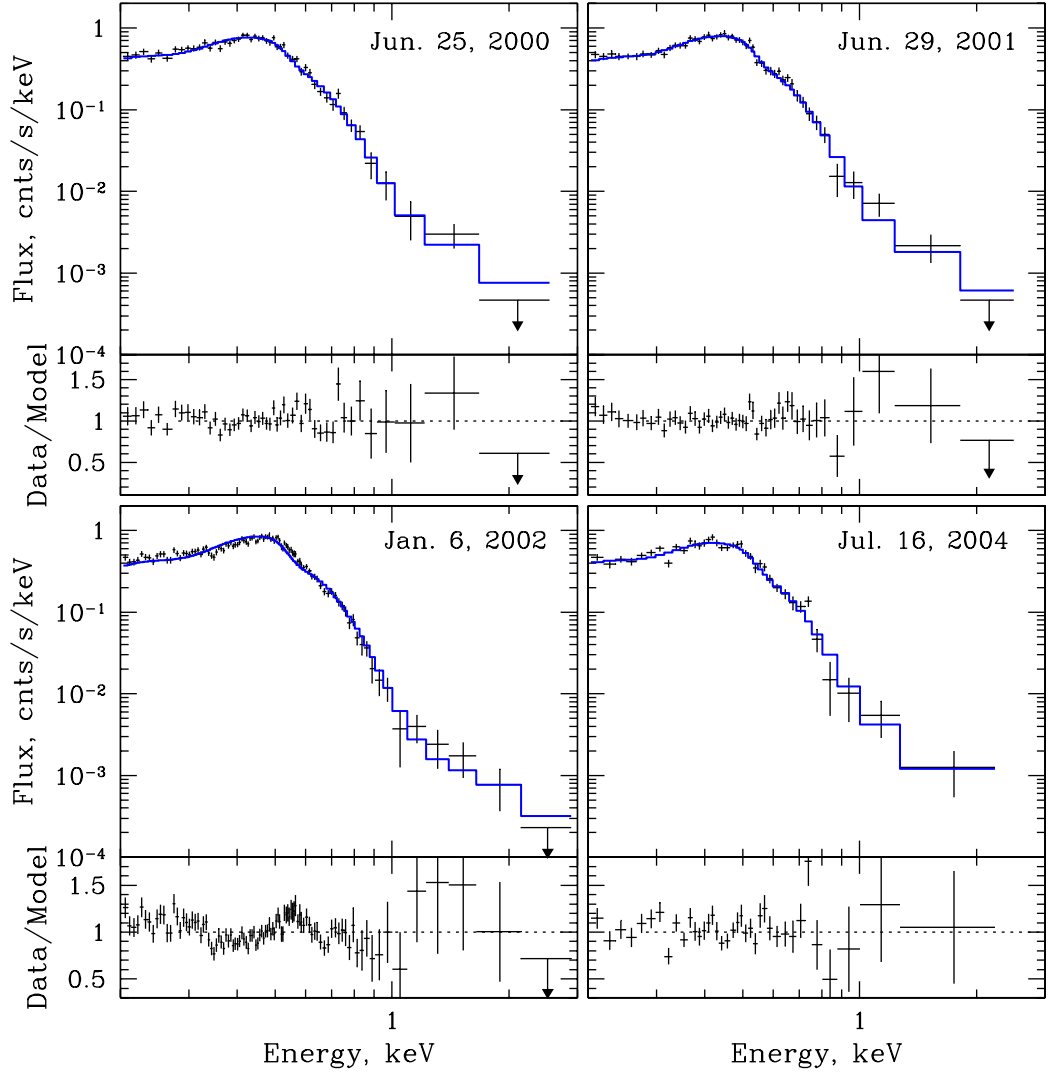}
\caption{EPIC-pn count spectra and model ratios of the source during during four {\em XMM-Newton} observations used in the 
analysis. The best-fit absorbed blackbody plus power law model approximation (Table \ref{timing_spec_par}) is shown with 
thick histograms. \label{spec_fig}}
\end{figure}

The pulse phase averaged spectra of XMMU J004252.5+411540 source are extremely soft, with only a small fraction of 
photons detected above 1 keV (Fig. \ref{spec_fig}). Nevertheless, the source spectra can not be adequately fit with 
a single soft spectral component (i.e. blackbody or other thermal models): an additional hard component is required 
to approximate the excess at higher energies (DiStefano et al. 2004; Orio 2006). We used a sum of the absorbed blackbody 
and power law models to approximate EPIC-pn and MOS spectra in the 0.2-3 keV energy range (Fig. \ref{spec_fig}). The 
best-fit spectral model parameters are shown in Table \ref{timing_spec_par}. The soft blackbody component has 
characteristic temperatures $kT\sim$62-67 eV and emitting radii $R_{bb}\sim 13000-20000$ km, and estimated bolometric 
luminosity $L_{bol}\sim (5-8)\times10^{38}$ ergs s$^{-1}$. The hard power law component is relatively weak, contributing 
less than 1\% of the total flux in the 0.2-1 keV energy range. The photon index of the power law component is poorly 
constrained for all four observations: the overall quality of the fit and soft component parameters are not particularly 
sensitive to the choice of the photon index. The equivalent absorbing column required by these model fits, 
$N_{\rm H}\sim (1.1-1.6)\times 10^{21}$ cm$^{-2}$ is somewhat higher than the expected foreground column in M31 
direction $7\times10^{20}$ cm$^{-2}$ (Dickey \& Lockman 1990). 

Although our simple two-component model provides good overall description of the source spectra, the relatively high 
reduced $\chi^{2}_{r}\sim$1.2-1.7 of the fits suggest a more structured shape of the source spectrum and a need for 
an additional components (see also Orio 2006). The approximation with the sum of absorbed blackbody and power law 
models leaves bump-like residuals in the 0.3-0.8 keV energy range, especially evident in the higher quality 2002 Jan. 
6 data (Fig. \ref{spec_fig}). These residuals can indicate the presence of discrete emission/absorption features 
intrinsic to the source, or be a result of incomplete spectral background subtraction (the source is embedded in the 
bright unresolved emission, that have a soft thermal spectrum rich in emission lines (Shirey et al. 2001)). 

To investigate the effect of additional discrete model features on the continuum spectral parameters, we added a 
Gaussian line to the existing two-component model, and used it to approximate EPIC-pn spectrum of 2002 Jan. 6 
observation. The addition of the Gaussian line at $\sim$0.53 keV greatly improved the quality of spectral fit 
($\chi^{2}=141.1$(143 dof) with $\Delta \chi^{2}\sim106$ for 3 additional parameters), and led to a significant 
change in the soft component temperature and emitting radius: $kT=76^{+3}_{-2}$ eV, $R_{bb}=6998^{+1131}_{-1391}$ 
km and an absorbing column depth, N$_{\rm H}=(9\pm1)\times10^{20}$ cm$^{-2}$. As a result, the estimated bolometric 
luminosity of the source dropped from $\sim7.6\times10^{38}$ ergs s$^{-1}$ to $\sim2.0\times10^{38}$ ergs s$^{-1}$. 
The application of this model to the data of 2000 Jun. 25 observation yields similar results: $kT=74^{+5}_{-3}$ eV, 
$R_{bb}=6602^{+1988}_{-1345}$ km, N$_{\rm H}=(8\pm2)\times10^{20}$ cm$^{-2}$ and bolometric luminosity of 
$\sim1.7\times10^{38}$ ergs s$^{-1}$ with $\chi^{2}_{r}=1.15$ for 111 dof. We also tried to add a set of the discrete 
absorption features (absorption lines and edges) to the original two-component model. This approach produced similar 
results: increase of the blackbody model temperature $kT\sim72-80$ eV and decrease of the emitting radius 
$R_{bb}\sim 7000-12000$ km, and overall decrease of the bolometric luminosity to $\sim(2-4)\times10^{38}$ ergs s$^{-1}$.

This is yet another demonstration of the huge uncertainty of SSS luminosity estimates, that has poorly understood 
systematic component determined by the quality of instrument calibration, background subtraction, foreground absorbing 
column estimation and choice of spectral model. This uncertainty extends well beyond standard statistical errors 
of the absorbed simple blackbody fits, usually adopted in the spectral analysis of SSSs, and can affect final 
conclusions about their properties. In the case of XMMU J004252.5+411540, a simple two-component model fits result 
in the estimated source luminosities that are super-Eddington assuming 1.4 M$_{\odot}$ compact object accreting 
hydrogen-rich material (DiStefano et al. 2004; Orio 2006, this work). Recently, Orio (2006) performed spectral 
analysis of {\em XMM}/EPIC-MOS data on XMMU J004252.5+411540 and obtained the lower limit on the source bolometric 
luminosity of $9.5\times10^{38}$ ergs s$^{-1}$ for 2001 June 29 and 2002 Jan. 6 observations, and suggested that the 
source could be linked to the ultraluminous supersoft X-ray sources found in other nearby galaxies and interpreted 
as IMBH candidates (Kong \& DiStefano 2005). Our analysis of the same datasets yields consistently lower source 
luminosity estimates ($\sim (5-8)\times10^{38}$ ergs s$^{-1}$) for the same model approximation, that become even 
lower ($\sim2\times10^{38}$ ergs s$^{-1}$) when a more complex models are used for spectral fitting.     

\section{DISCUSSION}
The X-ray pulsations and supersoft spectrum of XMMU J004252.5+411540 imply that it is almost certainly an accreting 
white dwarf in a binary system. The 217.7 s pulsation period is the shortest known among SSSs to date. It is too 
short to be interpreted as binary orbital period even for a double-degenerate system. The remaining possible 
explanations for the observed modulation include stellar rotation and non-radial $g$-mode pulsations of the white 
dwarf. Periodic variations on a time scales of 1000-2500 s have been recently detected in the postnova SSSs (Drake 
et al. 2003; Ness et al. 2003) and in one of the prototype SSS, CAL 83 (Schmidtke \& Cowley 2006), and interpreted 
as a signature of non-radial pulsations of the white dwarf. In the case of XMMU J004252.5+411540, the short period, 
long-term stability of the modulation and the absence of multi-periodicity argue against that interpretation. 
Therefore, it is more plausible to assume that the observed modulation results from rotation of a magnetized accreting 
white dwarf. The strength of the WD magnetic field should not be extremely high, so its spin and orbital periods are 
not locked. The energy spectrum of XMMU J004252.5+411540 is typical for luminous supersoft X-ray sources and soft 
intermediate polars. Assuming the source is located in M31, its observed luminosity ($\sim 3\times10^{37}$ ergs 
s$^{-1}$) is too high for intermediate polars (Haberl \& Motch 1995). On the other hand, the absence of a bright optical 
counterpart makes the interpretation of the source as Galactic foreground system unlikely. The remaining possibility 
is that XMMU J004252.5+411540 is a luminous supersoft source located in M31. In that case, XMMU J004252.5+411540 is 
the second M31 SSS with coherent X-ray pulsations detected, after the supersoft transient XMMU J004319.4+411758 
showing pulsations with 865.5 s period (Osborne et al. 2001; King et al. 2002).

The results of our spectral analysis suggest that the estimated source luminosity is probably close or may slightly 
exceed the isotropic Eddington luminosity for a typical white dwarfs. Assuming that the source luminosity 
($\sim2\times10^{38}$ ergs s$^{-1}$) is produced in the process of burning of the hydrogen-rich material on the WD 
surface, we can estimate the required accretion rate as $\dot{M} \sim 5\times10^{-7}$ M$_{\odot}$ year$^{-1}$ or 
$\sim3.2\times10^{19}$ g s$^{-1}$ (assuming the energy yield of nuclear burning 0.007$\dot{M}c^2$). Such a high 
accretion rate would imply either Roche lobe thermal time scale mass transfer from a near-main-sequence companion 
with a mass larger than that of the WD, or a mass transfer in a wide symbiotic (WD+red giant) system (van den Heuvel 
et al. 1992; Kahabka \& van den Heuvel 2006). 

If the X-ray pulsations observed in XMMU J004252.5+411540 result from rotation of a magnetized WD, accreting from 
the Keplerian disk (King et al. 2002), one can estimate the strength $B_{0}$ of the WD magnetic field. Assuming a 
dipole configuration of the pulsator magnetic field, the magnetospheric radius inside which the accretion flow is 
channeled by the field is (Davidson \& Ostriker 1973; Lamb, Pethick \& Pines 1973; Ghosh \& Lamb 1979a)
\begin{equation}
R_{\rm M}\sim0.5\left(\frac{\mu^{4}}{2GM_{\rm X}\dot{M}^{2}}\right )^{1/7} \simeq 8.5\times10^{9}\dot{M}^{-2/7}_{19}m^{-1/7}_{\rm X}\mu^{4/7}_{34} {\rm cm}
\end{equation}
where $\dot{M}$ denotes the accretion rate through the disk, $M_{\rm X}$ is a WD mass, $\mu=B_{0}R^{3}_{\rm wd}$ is 
the magnetic moment of the WD, $\dot{M}_{19}=\dot{M}/10^{19}$ g s$^{-1}$, $m_{\rm X}=M_{\rm X}/M_{\odot}$, and 
$\mu_{34}=\mu/10^{34}$ G cm$^{3}$. The matter will be magnetically channeled to the surface, if 
$R_{\rm M}\gtrsim R_{\rm wd}$. This condition is satisfied for a surface field 
\begin{equation}
B_{0}\gtrsim2.4\times10^{5}R_{9}^{-5/4}\dot{M}^{1/2}_{19}m^{1/4}_{\rm X}\,{\rm G} 
\end{equation}
where $R_{9}=R_{\rm wd}/10^{9}$ cm. For a $1M_{\odot}$ white dwarf with $R_{\rm wd}\sim10^{9}$ cm accreting matter at 
a rate $\dot{M}\sim3.2\times10^{19}$ g s$^{-1}$, $B_{0}\gtrsim4\times10^{5}$ G.     

In order to keep the ''propeller'' effect from preventing effective accretion onto white dwarf, the magnetospheric radius, 
$R_{\rm M}$ must be less than corotation radius $R_{\rm c}=(GM_{X}/4\pi^{2})^{1/3}P^{2/3}$, where $P$ is the WD spin 
period \cite{IS75}. As a result, we obtain

\begin{equation}
B_{0}\lesssim 8.6\times10^{3}R_{9}^{-3}\dot{M}^{1/2}_{19}m^{5/6}_{\rm X}P^{7/6}\,{\rm G}
\end{equation}
For the observed period $P=217.7$ s, $m_{\rm X}=1$, $R_{9}\sim 1$ and $\dot{M}_{19}\sim 3.2$, the estimated 
white dwarf surface magnetic field $B_{0}\lesssim8\times10^{6}$ G.

In the highly magnetized SSS scenario the accreted matter is channeled by the field and steadily arrives at the 
magnetic poles. Ideally, most of the hydrogen burning is expected to occur near the poles (King et al. 2002), 
creating bright hot spots. The observed energy dependence of the pulse profile supports this picture, suggesting 
a higher temperature of the emitting region responsible for the modulation. The estimated fractional polar cap 
area $f\sim (R_{\rm wd}/2R_{\rm M})\sim (R_{\rm wd}/2R_{\rm c})\sim 0.09$ (Davidson \& Ostriker 1973) implies a 
relatively large size of the accretion (hot spot) region, in agreement with quasi-sinusoidal pulse profile and 
low amplitude of the modulation ($\sim$10\%).

In the accretion process, angular momentum is transferred to the white dwarf and its rotation period should change 
(Pringle \& Rees 1972; Rappaport \& Joss 1977; Ghosh \& Lamb 1979b). Assuming that disk accretion is occurring in 
the system, and that all the specific angular momentum of the accreting material is transferred to the white dwarf 
at the magnetospheric radius $R_{\rm M}\lesssim R_{\rm c}$, one can estimate maximum rate of change of the spin 
period as
\begin{equation}  
\dot{P}\sim-7\times10^{-8}\dot{M}_{19}I^{-1}_{50}m^{2/3}_{\rm X}P^{7/3}\,{\rm s\,\,yr}^{-1}
\end{equation}
where $I_{50}$ is the moment of inertia of the white dwarf in units of $10^{50}$ g cm$^{2}$. For the typical white 
dwarf parameters ($I_{50}\sim 1$, $m_{\rm X}=1$), the observed spin period of $P=217.7$ s, and $\dot{M}_{19}\sim3.2$, 
the estimated spin-up rate is $|\dot{P}|\lesssim$0.065 s yr$^{-1}$. The period change of such magnitude could, in 
principle, be detected in XMMU J004252.5+411540 with sufficiently long ($>50$ ks) {\em XMM-Newton} observations 
spaced by a period of $\gtrsim 5$ years (Table \ref{timing_spec_par}). The measurement of the period change can be 
than used to further constrain the accretion rate in the system, providing an independent, better estimate of the 
bolometric luminosity of the source than presently possible with spectral analysis. Unfortunately, only one of the 
existing observations (2002 Jan. 6) is sufficiently long to obtain good estimate of the pulsation period. Future 
long {\em XMM-Newton} observations of the central region of M31 would allow a more detailed study of the evolution 
of the pulsation period of XMMU J004252.5+411540, and possibly improve our understanding of this and other luminous 
supersoft X-ray sources, both Galactic and extragalactic. 

\acknowledgements
The authors would like to thank France C\'ordova for careful reading of the manuscript and useful suggestions.Support for 
this work was provided through NASA Grant NAG5-12390. This research has made use of data obtained through the High 
Energy Astrophysics Science Archive Research Center Online Service, provided by the NASA/Goddard Space Flight Center. 
XMM-Newton is an ESA Science Mission with instruments and contributions directly funded by ESA Member states and the 
USA (NASA).

\clearpage

\begin{table}
\small
\caption{{\em XMM-Newton} Observations of M31 Used in the Analysis. 
\label{obslog}}
\small
\begin{tabular}{ccccccccl}
\hline
\hline
Obs.\# & Date, UT & Date, TJD  & Obs. ID  & Mission/Instrument & Mode/ & RA (J2000)\tablenotemark{a} & Dec (J2000)\tablenotemark{a} & Exp.(MOS/pn)\tablenotemark{b}\\
       &          &            &          &                    & Filter & (h:m:s)          & (d:m:s)           & (ks)      \\             
\hline
1 &2000 Jun. 25 &  & 0112570401 & {\em XMM-Newton}/EPIC& Full/Medium&00:42:43.00&41:15:46.1&28.9/24.9\\
2 &2001 Jun. 29 &  & 0109270101 & {\em XMM-Newton}/EPIC& Full/Medium&00:42:43.00&41:15:46.1&29.0/24.9\\
3 &2002 Jan. 06 &  & 0112570101 & {\em XMM-Newton}/EPIC& Full/Thin  &00:42:43.00&41:15:46.1&63.0/49.9\\
4 &2004 Jul. 16 &  & 0202230201 & {\em XMM-Newton}/EPIC& Full/Medium&00:42:42.12&41:16:57.1&19.6/16.4\\
\hline
\end{tabular}

\tablenotemark{a}{pointing coordinates}\\
\tablenotemark{b}{instrument exposure used in the analysis}\\
\end{table}

\clearpage
\begin{table}
\caption{X-ray Pulsation Parameters and Blackbody Spectral Fit Information for XMMU J004252.5+411540. 
\label{timing_spec_par}}
\small
\begin{tabular}{ccccccccccccl}
\hline
\hline
&\multicolumn{2}{c}{Timing Parameters}&\multicolumn{9}{c}{(BBODYRAD+POWERLAW)*WABS Spectral Model Parameters}\\
Obs. & Period &PF$_{0.2-1 keV}$      &N$_{\rm H}$&$kT_{\rm bb}$&$R_{\rm bb}$& Photon &Flux\tablenotemark{b}& $L_{\rm X}$\tablenotemark{c}&$L_{\rm X}$\tablenotemark{d}&$\chi^{2}$&Instrument\\
     &  (s)   & (\%)\tablenotemark{a}&($\times 10^{20}$ cm$^{-2}$)& (eV) & (km)& Index &             &      &     &  (d.o.f)   &        \\
\hline
1& $217.60(23)$ & $10.9\pm2.0$ & $12\pm1$ & $65^{+2}_{-1}$ & $14903^{+2001}_{-2157}$ & $2.59^{+1.20}_{-1.38}$ & $4.66\pm0.05$ & 3.22 & 3.1 & 135.8(114) & pn\\
 &              &              & $11\pm1$ & $67\pm2$       & $12912^{+1672}_{-1780}$ & $1.82^{+1.28}_{-1.30}$ & $5.02\pm0.06$ & 3.47 & 2.7 & 87.2(79) & MOS\\
2& $217.66(32)$ & $6.7\pm1.2$  & $14\pm1$ & $63\pm2$       & $18240^{+2572}_{-2755}$ & $2.63^{+1.14}_{-1.26}$ & $4.17\pm0.06$ & 2.88 & 3.8 & 133.1(114) & pn$^{f}$\\
 &              &              & $12\pm1$ & $63\pm2$       & $17737^{+2566}_{-2664}$ & $1.75^{+1.01}_{-1.39}$ & $5.01\pm0.08$ & 3.46 & 3.7 & 85.8(76) & MOS\\
3& $217.73(08)$ & $9.1\pm1.2$  & $16\pm1$ & $66\pm1$       & $19553^{+1644}_{-1858}$ & $1.75^{+0.67}_{-0.76}$ & $5.17\pm0.05$ & 3.57 & 5.5 & 247.3(146) & pn\\
 &              &              & $13\pm1$ & $63\pm1$       & $18462^{+1812}_{-2053}$ & $2.36^{+0.64}_{-0.69}$ & $4.78\pm0.06$ & 3.31 & 3.8 & 137.2(96) & MOS\\
4& $217.75(48)$ & $10.7\pm2.5$ & $12\pm1$ & $62^{+3}_{-2}$ & $17569^{+3529}_{-3211}$ & $3.06^{+1.65}_{-1.78}$ & $4.58\pm0.08$ & 3.16 & 3.4 & 134.5(100) & pn\\
\hline
\end{tabular}
\tablenotetext{a}{pulsed fraction in the $0.2-1$ keV energy band, defined as 
(I$_{\rm max}$-I$_{\rm min}$)/(I$_{\rm max}$+I$_{\rm min}$), where I$_{max}$ and I$_{min}$ 
represent source background-corrected intensities at the maximum and minimum of the pulse profile}
\tablenotetext{b}{absorbed model flux in the $0.2 - 1$ keV energy range in units of $10^{-13}$ 
erg s$^{-1}$ cm$^{-2}$}
\tablenotetext{c}{absorbed luminosity in the $0.2 - 1$ keV energy range in units of $10^{37}$ 
erg s$^{-1}$, assuming the distance of 760 kpc}
\tablenotetext{d}{estimated unabsorbed luminosity in the $0.2 - 1$ keV energy range in units of $10^{38}$ 
erg s$^{-1}$, assuming the distance of 760 kpc}
\tablenotetext{f}{part of the source flux falls into the gap between EPIC-pn CCDs}
\end{table}
\clearpage

\clearpage
\thispagestyle{empty}
\setlength{\voffset}{-25mm}
\clearpage
\setlength{\voffset}{0mm}

\end{document}